\documentstyle[11pt,titlepage]{article}  
\def\bm#1{\mbox{\boldmath $#1$}}
\def\Complex{\mbox{$I\!\!\!\!C$}} 
\titlepage  
 
\begin{document} 
\begin{center} 
{\Large \bf Transport along Null Curves} 
\vskip 1cm 
{\large Joseph Samuel and Rajaram Nityananda} \\ 
Raman Research Institute, Bangalore 560 080, INDIA.\\ 
\end{center} 
\vskip 2 cm 
\vskip 5 cm 
email:sam@rri.ernet.in, rajaram@rri.ernet.in 
\newpage  
\section*{Abstract} 

Fermi Transport is useful for describing the behaviour 
of spins or
gyroscopes following non-geodesic, timelike world 
lines. However, Fermi
Transport breaks down for null world lines. We introduce a transport law
for
polarisation vectors along non-geodesic null curves.  We show how this law
emerges naturally from the geometry of null directions by comparing
polarisation vectors associated with two distinct null directions. We then
give a spinorial treatment of this topic and make contact with the
geometric phase of quantum mechanics. There are two significant
differences between
the null and timelike cases. In the null case (i)  The transport law does
not approach a 
unique
smooth limit as the null curve approaches a null geodesic.
(ii)
The transport law for vectors is  
integrable,  
i.e the result depends only on the local properties of the
curve and not on the entire path taken. 
However, the transport of spinors is not integrable: 
there
is a global sign of topological origin.

\newpage 

\section{Introduction}
The spin four-vector $s^a$ of a
gyroscope (not acted on by external 
torques) moving along a timelike geodesic is parallel transported
along the geodesic\cite{Wein}. Similarly, the polarisation vector $v^a$
of a light ray following a null geodesic is parallel transported
along the null geodesic. 
If a gyroscope follows a timelike
curve which is not a geodesic, the parallel transport rule no 
longer applies: parallel transport does not preserve the
orthogonality of the spin four-vector to the tangent vector
for non-geodesic curves. The 
correct transport law for spin four-vectors along timelike
curves is Fermi transport\cite{Wein,MTW}. 
However, Fermi transport does not
apply to null curves: the transport equation breaks down. 
What is the appropriate transport law for polarisation vectors
along null, non-geodesic curves? The purpose of this paper is
to answer this question.

We will start in section II with a review of Fermi transport
and introduce in section III a new transport law for null, 
non-geodesic curves and describe its properties. 
We then show in section IV that this law derives naturally
from the geometry of null vectors just as Fermi transport derives
naturally from the geometry of timelike vectors. Section V is
a more sophisticated spinorial discussion of the 
transport law. Section VI connects  the 
spinorial discussion of Section V 
to  the geometric phase of a two state quantum system.
Section VII is a concluding discussion.

\section{Review of Fermi Transport} 
Let $({\cal M} ,g)$ be a space-time manifold with a 
Lorentzian metric $g$ of signature $(+,-,-,-)$. Let ${\cal C}$ be a smooth
time-like curve  in ${\cal M}$ and $p$ a point on the curve. 
In local coordinates,
the curve is described as $x^a(\tau)$, where $\tau$ is an arbitrary
parameter which increases into the future. We define the tangent
vector $\bf{t}$ by $t^a=dx^a/d\tau$ and the 
acceleration $\bf{\dot{t}}$ by $\dot{t}^a = t^b\nabla_bt^a$,
where $\nabla_b$ is the covariant derivative. 
If $s^a$ is a vector at $p$ which is orthogonal to $\bf t$ 
(${\bf s}\in T_p {\cal M}, 
({\bf s \cdot t})_p = 0$), Fermi transport 
gives us a vector at every point of ${\cal C}$
defined by the transport law
\begin{equation}
{Ds^a}/{d\tau} = F^{ab} s_b, \label{fermi}
\end{equation}

where

\begin{equation}
F^{ab} = \frac{(\dot{t}^a t^b - t^a\dot{t}^b)}{t^c\cdot t_c}, 
\label{deff}
\end{equation}

or more abstractly, ${\bf F = \dot{t}\wedge t/(t\cdot t)}$. The transport law
(\ref{fermi}) has the following properties:

\begin{enumerate}
\item Vectors orthogonal to the tangent vector at $p$ 
are transported to vectors 
orthogonal to the tangent vector at other points of ${\cal C}$.
\item The transport law is covariant under reparametrization of the 
curve. If $\tau$ is changed to $\tau^\prime$, where 
$f: = d\tau/d\tau^\prime > 0$, $\bf t$ 
and $\bf \dot{t}$ transform as follows
\begin{equation}
{\bf t} \rightarrow f{\bf t} \hspace*{1.5cm} \dot{{\bf t}} 
\rightarrow f^2\dot{{\bf t}} + \alpha_1{\bf t},
\end{equation}
where $\alpha_1$ is some function on ${\cal C}$. 

As a result ${\bf F} \rightarrow f {\bf F}$ 
and so the vector 
field defined on ${\cal C}$ by 
Fermi transport does not depend on the parametrization of 
${\cal C}$.

\item Inner products between vectors ${\bf s}_1$ and ${\bf s}_2$
are maintained
under Fermi transport
\begin{equation}
\frac{d({\bf s}_1\cdot {\bf s}_2)}{d\tau} = 
\frac{D ({\bf s}_1\cdot {\bf s}_2)}{d\tau} 
= F^{ab} s_{1a} s_{2b} + F^{ab} s_{1b} s_{2a} = 0
\label{inner}
\end{equation}
from the antisymmetry of $F^{ab}$. In particular,
 an orthonormal triad
$({\bf e}_1,{\bf e}_2,{\bf e}_3)_p$ of vectors orthogonal to ${\bf t}$ 
at $p$ can be Fermi transported along ${\cal C}$ to
give an orthonormal triad orthogonal to ${\bf t}$ everywhere on ${\cal C}$.

\item For geodesics Fermi transport reduces to 
parallel transport, since ${\bf F}$ vanishes on geodesics.
\end{enumerate}

Although the treatment given above is in the context of an 
arbitrary curved space-time, the notion of 
Fermi transport has nothing to do with curvature.This is evident
because the entire discussion takes place in the neighborhood
of a single open curve. By an appropriate choice
of co-ordinates, all the Christoffel symbols can be made to 
vanish on this curve \cite{LL}. The curvature of space-time thus plays
no essential role in the discussion of Fermi transport along an open
curve, which should
be viewed as a {\it special relativistic} kinematic effect. 
It is thus possible to conduct the whole discussion 
in flat space--time. By parallel transport we can identify the tangent 
space
at any point $p'$ along the curve ${\cal C}$ with $T_p{\cal M}$.
The subsequent discussion is hence entirely within $T_p{\cal M}$.

Fermi transport arises from the geometry of timelike
vectors in $T_p{\cal M}$. For a non--geodesic curve
the vector field obtained by  parallel transport of the tangent vector
at one point does not agree
with the local tangent vector at other points. Fermi transport allows us
to  set  up a 
correspondence between 
vectors orthogonal to distinct timelike vectors, viz the tangent
vectors at different points of a non-geodesic curve. This correspondence
is geometrically
natural but
is not integrable. By integrability we mean that the result of 
transporting a given initial vector to a final point along the curve
depens only on the local properties of the curve at the final point, and
not on the entire path upto that point. In this sense, the Fermi
transport rule is not integrable. This reflects the curvature of the space
of timelike directions, which
can be identified with the 
unit hyperboloid \cite{LL,Urbantke} in Minkowski space.

\section{Transport along Null Non-Geodesics}
As is evident from (\ref{deff}), the Fermi transport rule
breaks down for null curves,
since the denominator of (\ref{deff}) vanishes. 
For geodesic null curves, polarisation vectors are parallel
transported along the curve. 
But what about {\it non-geodesic} null curves? 
This is the question that we address and answer in this paper.
In this section we will simply write down a transport  law
for polarisation vectors along null, non--geodesic curves.
We will then show that this law has geometrically natural
properties as does Fermi  transport.

Let ${\cal N}$ be a smooth null curve described 
in local coordinates by $x^a(\tau)$,
where $\tau$ is an arbitrary parameter which increases into
the future. We use $l^a = dx^a/d\tau$ 
to denote the tangent vector to the null curve. 
We will also need $\dot{l}^a = l^b\nabla_b l^a$ 
and $\ddot{l}^a = l^b\nabla_b{\dot l}^a$. We will assume that
${\cal N}$ is nowhere geodesic i.e, ${\bf {\dot l}}$ 
is nowhere a scalar multiple of ${\bf l}$. By differentiating
${\bf l.l}=0$ with respect to $\tau$, 
we deduce in succession ${\bf l}.{\bf {\dot l}}=0$
and ${\bf {\dot l}}.{\bf {\dot l}}+{\bf l}.{\bf {\ddot l}}=0$. 
Let us write $H_p\subset T_p{\cal M}$ 
for the space of all vectors in 
$T_p{\cal M}$ which are orthogonal to ${\bf l}$ at $p$.
$H_p$ also includes ${\bf l}$,
since ${\bf l}$ is null. 
The Lorentz metric at $p$, pulled back to $H_p$, has signature
$(0,-,-)$. We need to ``mod out'' by the null direction 
$l^a$ to get a non-degenerate metric.
We define a polarization vector to be an equivalence class of
vectors in $H_p$ which differ by a multiple of ${\bf l}$ 
(${\bf v}_1 \sim {\bf v}_2 \iff {\bf v}_2 = {\bf v}_1 +
\lambda {\bf l}$ for some $\lambda$). 
Our transport law will tell us how to
transport polarization vectors along ${\cal N}$.
 (This ``modding out'' is standard for null curves \cite{HE}).

Given a polarisation vector (an equivalence class) at $p$,
let us take a representative element $v^a$ from $H_p$. 
We transport this
vector along $\cal N$ using the rule

\begin{equation}
\frac{Dv^a}{D\tau} = K^{ab} v_b, \label{Kattabomman}
\end{equation}
where $K^{ab}$ is an antisymmetric tensor defined by 

\begin{equation}
K^{ab} = \frac{(\ddot{l}^a\dot{l}^b - \dot{l}^a\ddot{l}^b)}
{(\dot{l^c}\cdot\dot{l_c})}. \label{defk}
\end{equation}
Finally we evaluate the equivalence class of $v^a$ (mod
out by $l^a$) to produce a polarisation vector
field along ${\cal N}$.

This law, designated $K$~transport from now on, 
has the following properties:
\begin{enumerate}
\item It is straightforward to check that polarization vectors
(equivalence classes 
of vectors perpendicular to ${\bf l}$)
at $p$ are transported to polarization vectors 
at other points of $\cal N$. 
\item The transport law is reparametrization covariant:
Under reparametrization of $\cal N$, 
${\bf l}\rightarrow f{\bf l}$ and 
\begin{equation}
\dot{{\bf l}}
\rightarrow f^2\dot{{\bf l}} + \alpha_2 {\bf l},\;\;\;
\ddot{{\bf l}}\rightarrow f^3\ddot{{\bf l}} + 
\alpha_3 \dot{{\bf l}} + \alpha_4 {\bf l},\;\;\;
\label{repar}
\end{equation}
where the $\alpha$'s are some functions on $\cal N$.
Using (\ref{repar}) in (\ref{Kattabomman},\ref{defk}), we can drop the $\alpha_2$ and $\alpha_4$ 
terms, which are proportional to ${\bf l}$: they either
vanish when they are contracted with $v_b$ in (\ref{Kattabomman})
or are modded out when we pass to polarisation vectors.
As a result,
\begin{equation}
\dot{{\bf l}}\cdot\dot{{\bf l}}\rightarrow f^4 
\dot{{\bf l}}\cdot\dot{{\bf l}}, \;\;\;\ddot{{\bf l}}\wedge
\dot{{\bf l}} \rightarrow (f^2\ddot{{\bf l}}+\alpha_3\dot{{\bf l}}) \wedge 
(f^3\dot{{\bf l}})=f^5\dot{{\bf l}}\wedge
\ddot{{\bf l}}.  
\label{dotdot}
\end{equation}
We find that 
\begin{equation}
K^{ab} \rightarrow f K^{ab}
\end{equation}
and so (\ref{Kattabomman}) provides a reparametrisation 
invariant transport law. 
\item Inner products between polarization vectors are preserved.
This follows from antisymmetry of $K$.

\end{enumerate}

A point worth noting is that $K$ transport involves 
the {\it second} 
derivative $\ddot{{\bf l}}$ of the tangent vector 
${\bf l}$. This is quite unlike 
Fermi transport which only involves 
the {\it first} derivative $\dot{{\bf t}}$
of the tangent vector. We will see below that this is an 
unavoidable consequence of the geometry of null vectors.

Unlike Fermi transport, the $K$ transport law does not have
a smooth limit as the null curve becomes a  geodesic. This will be
discussed in  the concluding section.

Note that we are not claiming to transport all vectors
along ${\cal N}$  in a geometrically natural manner. Our
rule is only meant for polarisation vectors, {\it i.e.} 
equivalence classes of 
vectors perpendicular to the tangent vector.

\section{Comparing Polarisation vectors on distinct Null vectors}
We will now show that the transport rule (\ref{Kattabomman}) originates
naturally from the geometry of null directions, just as the 
Fermi transport rule (\ref{fermi}) derives from the geometry
of timelike directions \cite{Urbantke}. If we parallel transport
the tangent vector ${\bf l}_{p'}$ 
along $\cal N$ from $p'$ to $p$, 
we find that for {\it non-geodesic} curves, the
parallel transported tangent vector 
does not agree with the local tangent vector ${\bf l}_p$.
The parallel tansport of a  vector orthogonal to the 
tangent vector at $p'$ 
is not in general
orthogonal to the local tangent vector ${\bf l}_p$, but
to the parallel transported
tangent vector. We need to find a way to compare vectors 
orthogonal to two distinct null vectors. We show below that
this can be done in a geometrically natural manner for
polarisation vectors.

Let $M$ be a four
dimensional vector space with Lorentzian metric ($\eta$) of signature
$(+,-,-,-)$.
($M$ is a model for $T_p{\cal M}$, the tangent space at the 
point $p$ of ${\cal M}$.) The set of future pointing ($l^0>0$),
null ($l^a l^b \eta_{ab} =0$) vectors in $M$ forms the future light cone
and the set of null directions (defined as future pointing 
null vectors modulo extent\cite{Penrose}) is a sphere.
We will sometimes refer to this sphere as the celestial 
sphere or the sky (although, strictly speaking, this
terminology should be reserved for {\it past pointing} null
directions). 

If $L$ is a null direction and ${\bf l}$ a null vector 
belonging to $L$, we define $H_L$  to be the space of vectors $v^a$ 
in $M$ orthogonal to $L$, $v^a l_a=0$. $H_L$ also includes
$L$. We define a polarisation vector ${\bf p}_L$ to be an equivalence
class of vectors in $H_L$ differing by an arbitrary multiple
of ${\bf l}$ ($v^a\sim v^a+\lambda l^a$ $\lambda$ arbitrary). 
The vector space of polarisation vectors defined by
$L$ is written $P_L$.

Given two distinct null directions 
$L_1$ and $L_2$  and a polarisation vector ${\bf p}_1\in P_{L_1}$,
there is a geometrically natural choice of a polarisation vector
from $P_{L_2}$. We pick from ${\bf p}_1$ the unique element 
${\bf w}\in H_{L_1}$
which is orthogonal to $L_2$ and define ${\bf p}_2$ as the class 
${\bf w}+\lambda_2 {\bf l}_2$ containing the vector ${\bf w}$.
More explicitly, pick any $\bm{v_1} \in {P_{L_1}}$.
Requiring that
\begin{equation}
({\bf v}_1 + \lambda_1 {\bf l}_1) \cdot {\bf l}_2 = 0
\end{equation}
uniquely fixes $\lambda_1$:
\begin{equation}
\lambda_1 = -\frac{{\bf v}_1 \cdot {\bf l}_2}{{\bf l}_1 
\cdot {\bf l}_2},
\end{equation}
which is well defined, since ${\bf l}_1 \cdot {\bf l}_2 > 0$,
for  ${\bf l}_1$ and ${\bf l}_2$  distinct. 
We write
\begin{equation}
{\bf w}= {\bf v}_1- \frac{{\bf v}_1 \cdot {\bf l}_2}
{{\bf l}_1\cdot {\bf l}_2} {\bf l}_1
\end{equation}
and define ${\bf p}_2$ to be the equivalence class
${\bf w} +{\lambda}_2 {\bf l}_2$ 
containing ${\bf w}$:
\begin{equation}
{\bf v}_2 = {\bf v}_1 -\frac{{\bf v}_1 
\cdot {\bf l}_2}{{\bf l}_1\cdot 
{\bf l}_2} {\bf l}_1 + {\lambda}_2 {\bf l}_2.
\end{equation}
Let us write $\delta {\bf l}_{12} = {\bf l}_2 - 
{\bf l}_1$ and note that ${\bf l}_2 \cdot {\bf l}_2 = 0
= ({\bf l}_1 + \delta {\bf l}_{12}) \cdot ({\bf l}_1 +
 \delta {\bf l}_{12}) = 2\delta {\bf l}_{12} \cdot
{\bf l}_1 + \delta {\bf l}_{12} \cdot \delta {\bf l}_{12}$ 
implies
\begin{equation}
{\bf l}_1 \cdot \delta {\bf l}_{12} = -\frac{1}{2} 
\delta {\bf l}_{12} \cdot \delta {\bf l}_{12}.
\end{equation}
We can now write 
\begin{equation}
{\bf v}_2 = {\bf v}_1 - \frac{2{\bf v}_1 
\cdot \delta {\bf l}_{12}}{\delta {\bf l}_{12} 
\cdot \delta {\bf l}_{12}} {\bf l}_1 + \lambda_2 {\bf l}_2
\end{equation}
and by suitable choice of ${\lambda}_2$
\begin{equation}
{\bf v}_2 = {\bf v}_1 -\frac{2({\bf v}_1 \cdot 
\delta {\bf l}_{12}) \delta {\bf l}_{12}}
{(\delta {\bf l}_{12} \cdot \delta {\bf l}_{12})}.
\label{rule}
\end{equation}
This  reverses the $\delta {\bf l}_{12}$ component of 
${\bf v}_1$ and leaves
the component of ${\bf v}_1$ which is orthogonal to 
$\delta {\bf l}_{12}$ unchanged.
The rule which associates polarisation vectors 
in $P_1$ to polarization
vectors in $P_2$  reverses orientation and 
therefore cannot be continuously
deformed to the identity. This can be seen 
quite clearly going to a 
frame in which ${\bf l}_1$ represents a light ray 
going in the positive $z$ direction 
and ${\bf l}_2$ a light ray going in the negative $z$ direction. 
Polarization vectors of ${\bf l}_1$ and ${\bf l}_2$ can then be identified
with the $x-y$ plane but
with opposite orientations. 

The rule for identifying $P_1$ and $P_2$ can 
also be stated succinctly as follows. 
Consider the two dimensional 
subspace of $M$ orthogonal to {\it both} ${\bf l}_1$ and 
${\bf l}_2$. The 
projector on to this subspace 
is $h_{ab} = \eta_{ab} - ({\bf l}_1\cdot{\bf l}_2)^{-1} 
(l_{1a} l_{2b} + l_{2a} l_{1b})$. This subspace 
can be identified with
{\it both} $P_1$ and $P_2$ (by taking its elements to 
represent classes) and this gives an identification of $P_1$ with
$P_2$. As the example of the last paragraph shows, 
the identification reverses orientation. The natural volume form
$\epsilon_{ab}:=({\bf l}_1\cdot {\bf l}_2)^{-1}
\epsilon_{abcd} l_1^c l_2^d$
on the two dimensional subspace reverses sign when 
${\bf l}_1$ and ${\bf l}_2$ are interchanged.

Since the map from $P_1$ to $P_2$ cannot be continuously deformed to the
identity, it does not have a smooth limit as $L_2$ tends to $L_1$. Such
a smooth limit would be necessary to define transport along a smooth null
curve. However, if we repeat the process by considering {\it three} null directions $L_1$, $L_2$, 
$L_3$ and go from $P_1$ to $P_2$ and $P_2$ to $P_3$ using the 
rule (\ref{rule}) twice, we get an orientation preserving map from $P_1$ to $P_3$.
This map does have a smooth limit as $P_1$, $P_2$ and $P_3$ approach
each other. It is intuitively clear that since we need {\it three} null
directions (rather than two) to take a smooth limit, the transport
law we derive for null vectors will depend on ${\bf l}, \dot{\bf l}$ and
$\ddot{\bf l}$, in contrast to the Fermi transport law, which only 
depends on ${\bf t}$ and $\dot{\bf t}$.
Writing ${\bf l}_1$, ${\bf l}_2$, 
${\bf l}_3$ for elements of 
$L_1$, $L_2$, $L_3$ and $\delta 
{\bf l}_{23} = ({\bf l}_2 - {\bf l}_3)$ we find as in (\ref{rule}),
\begin{equation}
{\bf v}_3 = {\bf v}_2 - \frac{2 {\bf v}_2 \cdot \delta {\bf l}_{23}}{\delta {\bf l}_{23} \cdot \delta {\bf l}_{23}} \delta {\bf l}_{23}
\end{equation}
using (\ref{rule}) we arrive at the following expression
for ${\bf v}_3 - {\bf v}_1$
\begin{equation}
\frac{4({\bf v}_1\cdot\delta {\bf l}_{12}) (\delta {\bf l}_{12}\cdot\delta {\bf l}_{23})\delta {\bf l}_{23} 
- 2({\bf v}_1\cdot\delta {\bf l}_{12})(\delta {\bf l}_{23}\cdot\delta {\bf l}_{23})\delta {\bf l}_{12}
-2({\bf v}_1\cdot\delta {\bf l}_{23})(\delta {\bf l}_{12}\cdot\delta {\bf l}_{12})\delta {\bf l}_{23}}
{(\delta {\bf l}_{12}\cdot\delta {\bf l}_{12})(\delta {\bf l}_{23}\cdot\delta {\bf l}_{23})}.
\label{express}
\end{equation}
We can now take the limit as the three null directions approach each other.
Let ${\bf l}(\tau )$ be the tangent vector of a smooth 
null curve. We expand ${\bf l}(\tau )$ in a Taylor series. 
\begin{eqnarray}
{\bf l}_1 &=& {\bf l}(\tau - \Delta\tau ) = 
{\bf l}(\tau ) -  \dot{{\bf l}}(\tau )\Delta\tau + 
1/2 \ddot{{\bf l}}(\tau ) (\Delta\tau )^2 + \ldots\\
{\bf l}_2 &=& {\bf l}(\tau )\\
{\bf l}_3 &=& {\bf l}(\tau + \Delta\tau ) = 
{\bf l}(\tau ) + \tau \dot{{\bf l}}(\tau )\Delta + 
1/2 \ddot{{\bf l}} (\Delta\tau )^2 + \ldots
\end{eqnarray}
where the ellipsis stand for higher order terms than we need. As a result,
\begin{eqnarray}
\delta {\bf l}_{12} &=& \dot{{\bf l}}(\tau )\Delta \tau - 1/2 \ddot{{\bf l}}(\Delta\tau)^2 
+ \ldots\\
\delta {\bf l}_{23} &=& \dot{{\bf l}}(\tau )\Delta \tau + 1/2 \ddot{{\bf l}}(\Delta\tau)^2 + \ldots
\end{eqnarray}
The leading term in the denominator of 
(\ref{express}) is of order $(\Delta\tau )^4$:
\begin{equation}
(\dot{{\bf l} }(\tau )\cdot\dot{{\bf l} }(\tau ))(\Delta\tau )^4.
\end{equation}
In the numerator the term of order $(\Delta\tau )^4$ is 
\begin{equation}
[4({\bf v}_1\cdot\dot{{\bf l} })(\dot{{\bf l} }
\cdot\dot{{\bf l} })
\dot{{\bf l} }-2({\bf v}_1\cdot\dot{{\bf l} })(\dot{{\bf l} }
\cdot\dot{{\bf l} })\dot{{\bf l} } - 2({\bf v}_1\cdot\dot{{\bf l} }) 
(\dot{{\bf l} }\cdot\dot{{\bf l} })\dot{{\bf l} }  
]
(\Delta\tau )^4
\end{equation}
which vanishes. The first non vanishing term is of order $(\Delta\tau )^5$. 
After some straight forward algebra, we evaluate 
$({\bf v}_3-{\bf v}_1)/(2\Delta\tau)$
and find that the limit $\Delta\tau\rightarrow 0$ exists and 
yields the $K$ transport law (\ref{Kattabomman}) of 
 section II.

As the reader may have noticed, the entire 
discussion of this section depends only on the conformal 
metric and not the metric itself. The definition of $H_p$
and ``modding out'' by ${\bf l}$ are unchanged under
conformal transformations and so is the rule (\ref{rule}) for
comparing polarisation vectors between fibres. One may therefore
expect that the $K$ transport rule is conformally invariant. 
It is easily checked that it is. The $K$ transport of a vector
using a conformally rescaled metric only results in trivial
rescalings of the polarisation vector (to be expected because
parallel transport preserves the norm). 
Under conformal transformations there is no change in the 
direction of polarisation of the $K$ transported vector.

\section{Spinorial formulation}
As one might expect, the discussion of the last section can be
formulated quite naturally in terms of spinors \cite{Penrose}. Let
$(V,\epsilon_{AB})$ be a complex 2-dimensional vector space 
endowed with an antisymmetric non-degenerate tensor $\epsilon_{AB}$.
Elements of $V$ are written $\xi^A$. The complex conjugate of $\xi^A$
is written $\bar{\xi}^{A'}$, with $A'$ a ``primed'' or ``dotted''
spinor index. A pair $AA'$ of spinor indices 
can be converted into a vector index $a$ by using the standard
correspondence between vectors and spinors:
\begin{equation}
v^a=\sigma^a_{AA'} v^{AA'},
\label{correspond}
\end{equation}
The components of $\sigma^a$ are $\sigma^0=I$,
$\sigma^1=\sigma^x$,$\sigma^2=\sigma^y$,$\sigma^3=\sigma^z$,
and $I$ is the $2\times 2$ identity matrix and 
$(\sigma^x,\sigma^y,\sigma^z)$ are the standard Pauli matrices.
We will write such relations (\ref{correspond}) as 
$v^a\rightleftharpoons v^{AA'}$.The spinor
$\xi^A$ defines a future pointing null 
vector ${\bf l}^a \rightleftharpoons \bar{\xi}^{A'} \xi^A$.
Altering $\xi^A$ by a phase does not alter the vector all and multiplying
$\xi^A$ by a real number alters the {\it extent} of the null vector,
but not its direction.

 Define the following equivalence
relation on (non-zero elements of) $V$:
\begin{equation}
\xi^A \sim \alpha\xi^A,
\end{equation}
where $\alpha$ is any non zero complex number. The set of equivalence classes
form a sphere $S^2$, which is the set of future pointing null 
directions - the sky of the previous section. 
Non zero elements of $V$ form a fibre bundle 
with the base equal to $S^2$  and the fibre isomorphic to the
set of non zero complex numbers. The phase of this non-zero
complex number determines a ``flag plane'' \cite{Penrose}
or polarisation direction.  This is easily seen as follows.
Let $L_1$ be a point on $S^2$ and $\xi^A_1$ a point on the fibre
over $L_1$. The null vector $l^a 
\rightleftharpoons \bar{\xi}_{1}^{A'}
\xi^A_1$ belongs to the null direction $L_1$. 
Let us pick an arbitrary spinor
$\xi_2$, distinct from $\xi_1$, so that $\xi_{2A}\xi_1^A$ is nonzero. 
One can always multiply it by a suitable complex number so that
$\xi^A_1 \cdot \xi_{2A} = 1$. The space-like unit vector
\begin{equation}
v^a \rightleftharpoons \frac{1}{\sqrt{2}}
(\bar{\xi}_{1}^{A'} \xi^A_2 + \bar{\xi}_{2}^{A'} \xi^A_1)
\end{equation}
is clearly orthogonal to $l_{1a} \rightleftharpoons \bar{\xi}_{1}^{A'}
\xi^A_1$
\begin{equation}
l_{1a}v^a = \frac{1}{\sqrt{2}}\bar{\xi}_{1A'} \xi_{1A} (\bar{\xi}_{1}^{A'} \xi^A_2 +
\bar{\xi}_{2}^{A'} \xi^A_1) = 0
\end{equation}
A different choiceof $\xi_2$, obtained by adding a multiple of $\xi_1$ to
it,  only changes the 
vector $v^a$ in (28)  by a multiple of $l^{a}_{1}$. Thus $\xi^{A}_{1}$ 
determines an
equivalence class of unit vectors $v^a$ orthogonal to 
$l_a$, i.e., a unit
polarization vector.
As can be easily verified, altering $\xi^{A}_{1}$ by a phase 
$e^{i\theta}$
leads to a rotation of the polarization vector by an 
angle $2\theta$. Thus, 
$\xi^{A}_{1}$ and $-\xi^{A}_{1}$ define the same 
polarization vector.
The correspondence 
is two to one.  
We thus have a map from the  fibre ${\cal F}(L)$ over a 
null direction $L$ to the unit circle of 
polarization vectors defined by $L$. Our discussion
now will be entirely on the spinor bundle.

Given two distinct points $L_1$ and $L_2$ on the base and a point
$\xi^{A}_{1}$ on the fibre ${\cal F}(L_1)$ over $L_1$, there is a natural
way to pick a point $\xi^{A}_{2}$ the ${\cal F}(L_2)$ fibre over $L_2$. 
We pick the unique point $\xi^{A}_{2}$ which satisfies
\begin{equation}
\xi^{A}_{1}  \xi_{2A} = 1. \label{map}
\end{equation}
(This choice when translated into vectors agrees with the discussion of
section IV). The rule (\ref{map}) is well defined only if $L_1$ and
$L_2$ are {\it distinct} points. If $\xi^{A}_{1}$ is altered by a phase
$\xi^{A}_{1} \rightarrow e^{i\theta} \xi^{A}_{1}$, $\xi_2^A$ picks 
up the opposite phase: $\xi^{A}_{2} \rightarrow
e^{-i\theta} \xi^{A}_{2}$. Thus, rule (\ref{map}) maps a circle
winding in the anticlockwise sense to a circle winding in the clockwise
sense. The map defined by 
(\ref{map}) from ${\cal F}(L_1)$ to 
${\cal F}(L_2)$ cannot be continuously deformed 
to the identity and  the
rule (\ref{map}) does not admit a smooth 
limit as $L_2$ approaches $L_1$.

As in Section 4, we can solve this problem by considering {\it three}
points, $L_1$, $L_2$,
$L_3$ on $S^2$. Given $\xi^{A}_{1} \epsilon {\cal F}(L_1)$ we pick $\xi^{A}_{2}$
from ${\cal F}(L_2)$ accordingly to the rule (\ref{map}) and 
repeat the process to pick
$\xi^{A}_{3}$ from $L_3$ using $\xi^{A}_{2}  \xi_{3A} = 1$. 
The map from
$L_1$ to $L_3$ does admit a smooth limit as $L_1,L_2$ and $L_3$
approach each other. We will use this below to derive the
spinorial form of the transport law (\ref{Kattabomman}).
If $L_1,L_2$ and $L_3$ are three distinct null directions, 
the map from $L_1$ to $L_3$ (via $L_2$) does depend on $L_2$.
If a different choice $L_2'$ is made, one can check that the point $\xi_3$ 
on ${\cal F}(L_3)$ determined by $\xi_1$ is multiplied by
a complex number $\chi$, where
\begin{equation}
\chi=\frac{(\xi_1^A\xi_{2'A})(\xi_2^B\xi_{3B})}
{(\xi^{D}_{1}\xi_{2D})(\xi_{2'}^C\xi_{3C})}.
\label{cross}
\end{equation}
$\chi$ depends only on the four null directions $L_1,L_2,L_3$
and $L_{2'}$ and not on the representatives chosen from each
fibre. $\chi$ is called the cross ratio \cite{Penrose} of these
four null directions.
The fact that the map from $L_1$ to $L_3$ {\it does} depend
on $L_2$, ($\chi$ is not the identity) shows that the {\it discrete}
rule for comparing fibres over distinct null directions
is not integrable.

We will now take the continuous limit of the discrete rule and 
recover the transport law (\ref{Kattabomman}) in spinorial language.
We are given a curve $L(\tau)$ of null directions and a point
$\xi(0)$ on the fibre over $L(0)$. What we seek 
is a geometrically natural ``lift'' of this curve
{\it i.e.} we need to find $\xi(\tau)$ so that $\xi(\tau)\in 
{\cal F}(L(\tau))$.

Fix a spin frame $(\dot{i}^A,o^A)  
(i^Ao_A = 1)$ and write
$\xi(\tau ) = \gamma(\tau ) 
(i^A + z(\tau )o^A)$. $z$ is a stereographic
coordinate on the set of null directions and $\gamma$ 
is a coordinate on the 
fibre. The problem now is: give a smooth 
curve $z(\tau)$, determine $\gamma(\tau)$ using the rule 
(\ref{rule}). We expand $z(\tau)$in a Taylor series and write
\begin{eqnarray}
z_1 &=& z(\tau-\Delta\tau) = z(\tau) - \dot{z}\Delta\tau + 
\frac{1}{2} 
\ddot{z}(\Delta\tau)^2 + \ldots\\
z_2 &=& z(\tau)\\
z_3 &=& z(\tau+\Delta\tau) = z(\tau) + \dot{z}\Delta\tau + 
\frac{1}{2} \ddot{z}(\Delta\tau)^2.
\end{eqnarray}
Evidently,
\begin{eqnarray}
\xi_1 &=& \xi(\tau +\Delta\tau ) = \gamma (\tau -\Delta\tau ) 
(i^A+z(\tau-\Delta\tau )o^A) \\
\xi_2 &=& \xi(\tau ) = \gamma (\tau )
(i^A+z(\tau)o^A)\\
\xi_3 &=& \xi(\tau+\Delta\tau ) = 
\gamma (\tau\Delta+\tau )(i^A+z(\tau +\Delta\tau)o^A).
\end{eqnarray} 
Using  the rule (\ref{rule}) for determining 
$\xi_2$ and $\xi_3$ we find from 
$\xi^A_1 \xi_{2A} = \xi^A_2 \xi_{3A} = 1$,
\begin{eqnarray}
\gamma (\tau )\gamma (\tau -\Delta\tau ) (z(\tau ) - z(\tau -\Delta\tau ) &=& 1\\
\gamma (\tau )\gamma (\tau +\Delta\tau ) (z(\tau +\Delta\tau ) - z(\tau ) &=& 1.
\end{eqnarray}
We eliminate $\gamma (\tau )$ from these equations 
and find using the Taylor 
expansion for $z(\tau )$
\begin{equation}
\frac{\gamma (\tau +\Delta\tau ) - \gamma (\tau -\Delta\tau)}
{\gamma (\tau -\Delta\tau )} 
= \frac{(\dot{z}\Delta\tau - 1/2 \ddot{z}(\Delta\tau )^2 + \ldots) - 
(\dot{z}\Delta\tau + 1/2 \ddot{z}(\Delta\tau)^2 + \ldots)}
{(\dot{z}\Delta\tau + \ldots )}.
\end{equation}
Evaluating $$\frac{\gamma (\tau +\Delta\tau ) -
 \gamma (\tau -\Delta\tau )}{\gamma (\tau -\Delta\tau ) 2\Delta \tau}$$ 
 we find in the limit $\Delta\tau \rightarrow 0$
\begin{equation}
\gamma ^{-1} \dot{\gamma } = (-1/2) \ddot{z}/\dot{z} \label{Kspinor}
\end{equation}
which is the transport law (\ref{Kattabomman}) in spinorial form. Its this form, it is apparent
that (\ref{Kspinor}) can be integrated to yield 
\begin{equation}
\gamma(\tau)=\gamma(0) \sqrt{(\dot{z}(0)/\dot{z}(\tau))}.
\label{int}
\end{equation} 
Although the discrete rule (\ref{map}) for comparing
points on distinct fibres is not integrable, its continuous
limit (the K-transport rule) is. This is a feature of K--transport
which is not shared by Fermi transport. It is interesting that the
discrete rule is not integrable, while the continuum limit is. This
indicates that as the four null directions $L_1, L_2,L_2',L_3$
tend to each other, the phase discrepancy between alternative paths
vanishes sufficiently fast that there is none left in the continuum limit.

\section{Relation to the Geometric Phase in two state quantum mechanics}
 The discussion so far has been entirely Lorentz invariant.
 In particular the last section has been in the language of
 $SL(2,\Complex)$ spinors. We will now attempt to make contact
 with the notion of  the Geometric Phase in quantum mechanics \cite
{Berry}. The motivation is as follows. One can think of the spinors of the
previous section as describing the state vectors of a  two state quantum
mechanical system. The
equivalence relation (27) defining the fibres is precisely the one  which
takes one from Hilbert 
space to the space of physical states (ray space). The relation (30)
associates a  member $\xi_2$ of the fibre over
the point $L_2$ is associated with a particular member $\xi_1$ of the
fibre over $L_1$. In quantum mechanics, there is  a
notion of two state vectors, corresponding to different rays, being "in
phase", which  leads to the geometric phase.
It is natural, therefore, to ask whether the relation (30) has
an analogue in  quantum mechanics. At first
sight, there is an obstacle. The relation (30) breaks
down
when the two null vectors $L_1$ and $L_2$ are {\it coincident}. The
geometric phase
convention in quantum mechanics breaks down when the two rays are {\it
orthogonal}. Nevertheless, there is a close correspondence which is
developed in this section. 
 In order to 
 do this we need to reduce the structure group 
 from $SL(2,\Complex)$ to $SU(2)$.  
 The $SL(2,\Complex)$ 
 invariant structures that we described in the last section will
 now be described in  terms of $SU(2)$ spinors.

 In order to break the structure group down from 
 $SL(2,\Complex)$  to $SU(2)$, we introduce 
 on the two
 complex dimensional vector space $(V,\epsilon_{AB})$
 an additional structure: a positive definite
 Hermitian inner product $G_{A'A}$. One can think of $G_{AA'}$ as the
spinor corresponding to a timelike four vector. Thus making a choice of
$G$ is like making a choice of the four-velocity of a frame of reference,
which still leaves freedom to make spatial rotations.  The group of
 transformations that preserves both $\epsilon_{AB}$ and 
 $G_{AA'}$ is $SU(2)$.  By choice of spin frame 
 $(\iota^A,o^A)$ we can 
 arrange that
 \begin{equation}
 G_{A'A}=\iota_A\iota_{A'}+o_A o_{A'}
\end{equation}
and use $G_{A'}{}^A$ to define a $\dagger$ operation taking  
a spinor $\xi^A$ to a new spinor $\xi^{\dagger A}$ transforming in the
same way.
\begin{equation}
\xi^{\dagger A}:=\overline{\xi}^{A'} G_{A'}{}^A
\end{equation}
 We will sometimes use Dirac notation $|\xi >$ for 
 the element 
 $\xi^A$ of 
 $V$ and $<\xi| $
 for the element $\xi^\dagger_A$ of $V^*$ (the dual of $V$). 
 Note that $<\xi |$ is not 
 $\xi_A$, for $<\xi |\xi >$ is positive definite 
 whereas $\xi_A\xi^A$ vanishes. It is
 easily checked that $\xi^{\dagger\dagger}_A = -\xi_A$ 
 and that $\xi^\dagger$
 is orthogonal to $\xi, i.e  <\xi^\dagger |\xi > = 0$. 
 The action of $\dagger$ on the sphere of null directions 
 (the sky) is easy to visualise. By 
 explicit computation are sees that $\iota^\dagger = o$, 
 $o^\dagger = -\iota$ and 
 so, if $\xi^A = i^A + z o^A$, 
 $ \xi^\dagger{}^A = -\bar{z}i^A + o^A$, 
 so $\dagger$ sends each point on
 the sky to its antipode. The subgroup of $SL(2,\Complex)$ which 
 preserves the relation of antipodality is
 $SU(2)$, which acts on the sky by rotations.

We now  identify the $SU(2)$ spinors with state vectors of a two state
system 
and the sky with the corresponding ray space, which is a sphere. 
(Historically, this arose in the context of polarised light, which can be
represented by a pair of complex numbers, and the sphere was discovered by
Poincar{\'e} after whom it is named. The definition that two states are in
phase when their  inner product is real and positive was proposed by
Pancharatnam \cite {Berry}). The rule $\xi^A_1\cdot
\xi_{2A} = 1$ for comparing points on distinct fibres 
can be rewritten as 
\begin{equation}
\xi^{\dagger \dagger}_{1A}\xi_2^A=<\xi^\dagger_1|\xi_2>=1
\label{Prule}
\end{equation}
i.e, we require that $ |\xi_2>$ be {\it in phase}  
with $|\xi^\dagger_1> $. (We are not concerned here
with the modulus of the complex number 
$ <\xi^\dagger_1|\xi_2>$, but only its phase.
) The rule (\ref{Prule}) is well defined 
if $\xi_1 $ and $\xi_2 $ are on
distinct fibres, or, equivalently, if  $\xi_2$ and $\xi^{\dagger}_1$
are not antipodal. 

As in the case of vectors,  the rule (30) for passing
from the fibre  ${\cal F}(L_1) $ over  $ L_1$ to the 
fibre $ {\cal F}(L_3)$ over $L_3$ 
(via $L_2$ ) will depend on $L_2$. 
This dependence is captured
by the cross ratio (\ref{cross}). The phase of the complex
number (\ref{cross}) is a measure of the non-integrable 
nature of the rule (30) for comparing points on distinct fibres.
We can rewrite this quantity as the phase of the complex
number 
\begin{equation}
<\xi_1|\xi^{\dagger}_2><\xi^{\dagger}_2|\xi_3>
<\xi_3|\xi^{\dagger}_{2'}><\xi^{\dagger}_{2'}|\xi_1>,
\label{Panch}
\end{equation}
which has a simple geometric interpretation. Consider 
the four points  $ L_1,\tilde{L_2},\tilde{L_{2'}} $ and $L_3$ 
on the celestial sphere, where $ \tilde{L_2}$ and 
$\tilde{L_{2'}}$ 
are points antipodal to $L_2$ and $L_{2'}$ 
respectively. The phase of
$\chi$ has an interpretation which is well known
in the  
Geometric phase literature \cite{Berry}: it is equal to half the solid
angle subtended at the center of the sphere by the geodesic
rectangle $  L_1,\tilde{L_2}, L_3,\tilde{L_{2'}},L_1$. It follows 
that the change in the plane of polarisation in following
the route $ L_1,\tilde{L_2}, L_3, \tilde{L_{2'}},L_1$ is equal to
the solid angle subtended by this rectangle. Although this change in the
plane of polarisation has been computed in language pertaining to a given
frame of reference, it is of course Lorentz invariant, from the earlier
discussion.

\section{Concluding Discussion}

We have presented a transport law (\ref{Kattabomman}) which 
is the replacement for  Fermi
transport in the case of  null curves. We have also shown 
how this transport law arises
naturally from the geometry of null vectors. The $K$ transport
law has a natural description in terms of $SL(2,\Complex)$ spinors.
This description also brings out close analogies with the geometric
phase, once it is specialised to $SU(2)$ spinors by choosing a timelike
observer. In the rest of this section we compare the $K$ transport law
and Fermi transport.

The main difference between 
Fermi transport and $K$ transport is due to the difference
between the geometry of timelike directions and the geometry
of null directions. The set of timelike directions can be 
identified with a time-like 3-hyperboloid, whose isometry group
is the entire Lorentz group. In contrast, the null directions
are identified with a 2-sphere and the Lorentz group acts on the 
sphere by {\it conformal} transformations. There is consequently
no Lorentz invariant meaning to the statment that two null
directions are ``near'' each other. By a suitable Lorentz
transformation, any two distinct null directions can be made
antipodal. As a result, 
there is no Lorentz and reparametrization invariant measure
of the ``acceleration'' of a null curve. If the direction 
of the tangent vector of a null curve changes ``slightly''
in one Lorentz frame, this deviation
can be made as large as one pleases in some other
Lorentz frame. 

Fermi transport reduces smoothly to parallel transport when the timelike
curve becomes a timelike geodesic. In contrast,  
the transition from null curves to null geodesics
is a singular one. This is reflected in the 
absence of a smooth limit for $K$ transport. 
As an example, let $\cal M$ 
be Minkowski space with standard $(t,x,y,z)$ Cartesian 
co-ordinates and consider the null curve: 
$x=R \cos(\Omega \tau),
y=R\sin(\Omega \tau), z=\tau, t=(\sqrt{1+R^2\Omega^2}) \tau$, 
where $R$ and $\Omega$ are constants. This curve describes a particle
moving at the speed of light along a helical path. 
 
The $K$ tensor for this 
null curve is easily worked out to be $K=\Omega dx\wedge dy$
and is independent of  $R$. The transport rule
simply says that the polarisation vector rotates about the $z$-axis with
angular velocity $\Omega$. This is of course the angular velocity of the
frame made up by the spatial tangent and the normal, i.e the Serret-Frenet
frame.
 
In the limit that $R$ tends to $0$ with $\Omega$ finite, the null curve
does become a geodesic curve. However, $K$-transport does not reduce to
parallel transport Thus, the limit of $K$ transport to null geodesics is
not smooth.  In fact, one can easily do the above Minkowski-space
calculation for motion at the speed of light along a general smooth space
curve whose spatial curvature nowhere vanishes. We can choose the length
of the curve as a parameter, and choose polarisation vectors to be purely
spatial. As in the case of the helix, one finds that $K$ transport reduces
to Serret-Frenet transport, which is an integrable rule. One other way to
check
this is to compute the rate of change of the 
cosine of the angle made by
the transported vector ${\bf v}$ with 
the acceleration vector $\dot{\bf
l}$. A straightforward calculation using $K$--transport
shows that $$\frac{d}{d\tau}(\frac{
{\bf
v}\cdot {\dot {\bf l}}} {\sqrt{{\dot {\bf l}} \cdot
{\dot {\bf l}}}})=0$$ 
Since the acceleration 
vector vanishes for geodesics, the rule must and does become ill-defined
in the geodesic limit, as does Serret-Frenet transport.

Although the discrete rule given above for comparing polarisation vectors
on distinct fibres is not integrable, the limit of this rule for smooth
curves {\it is} integrable. This is a feature of the $K$ transport rule,
which is different from Fermi transport. We note that if the rule had not
been integrable in the continuous limit, we would have been able to define
a two-form on the base space (the sky) whose integral around a closed
curve would give the total rotation on traversing that curve. But, as is
well known, there is no Lorentz invariant notion of area on the space of
null vectors, except the one which assigns zero to each such area. 
With hindsight, the integrability is consistent with, and even
forced by, Lorentz invariance.

Finally, we close with the remark that on spinors,
$K$ transport is only {\it locally} integrable. (\ref{int})
fixes $\gamma(\tau)$ only up to a sign. This sign is
unimportant in discussing the transport of polarisation
vectors. However, in transporting spinors along null, 
non--geodesic curves one can see a non integrable phase
of topological origin with values $\pm1$. 
The result of spinor transport is not affected by 
continuous deformations of  
the null curve $\cal N$  
within  the class of null, everywhere 
non--geodesic curves, connecting the 
end points of $\cal N$,
but {\it is} affected by
changes  in $\cal N$ which cannot be continuously deformed away. 
Let $z(\tau)$
be a smooth closed simple (i.e non self-intersecting) curve in the space
of null directions,
with ${\dot z}(\tau)$ being nowhere zero. The complex number  ${\dot z}
(\tau)$
encircles the origin once in the complex plane for such a curve. Equation
(42) then shows that transporting
a spinor once along  this curve results in a phase difference
of $\pi$, which is in principle observable by interference.
{\it Acknowledgments:} J.S. thanks V.P. Kattabomman for 
a discussion on transport rules.

\end{document}